\documentclass[twocolumn,showpacs,preprintnumbers,amsmath,amssymb]{revtex4}

\usepackage{graphicx}
\usepackage{dcolumn}
\usepackage{bm}

\begin{document}

\preprint{Accepted for Phys. Rev. B (2011)}

\title{Aharonov-Bohm Exciton Absorption Splitting in Chiral Specific Single-Walled Carbon Nanotubes in Magnetic Fields of up to 78 T}

\author{Shojiro Takeyama}
 \email{takeyama@issp.u-tokyo.ac.jp}
\author{Hirofumi Suzuki}%
\affiliation{Institute for Solid State Physics, The University of Tokyo, 5-1-5, Kashiwanoha, 
Kashiwa, Chiba, 277-8581, Japan.}
%
\author{Hiroyuki Yokoi}
\affiliation{Department of Materials Science and Technology, Kumamoto University, Kurokami, Kumamoto 860-8555, Japan}
\author{Yoichi Murakami}
\affiliation{Global Edge Institute, Tokyo Institute of Technology, Ookayama, 152-8550, Japann}
\author{Shigeo Maruyama}
\affiliation{Department of Mechanical Engineering, School of Engineering, The University of Tokyo, Hongo, 113-8656, Japan}

\date{\today}

\begin{abstract}
The Ajiki-Ando (A-A) splitting of single-walled carbon nanotubes(SWNT) originating from the Aharanov-Bohm effect was observed in chiral specific SWNTs by the magneto-absorption measurements conducted at magnetic fields of up to 78 T. The  absorption spectra from each chirality showed clear A-A splitting of the $E_{11}$ optical excitonic transitions. The parameters of both the dark-bright exciton energy splitting and the rate of A-A splitting in a magnetic field were determined for the first time from the well-resolved absorption spectra. 
\end{abstract}

\maketitle

\section{\label{sec:level1}Introduction}

A single-walled carbon nanotube (SWNT) is one of an ideal nanosystem for observing the splitting of energy bands upon application of an external magnetic field oriented parallel to the tube axis. This is known as the Aharanov-Bohm effect. The splitting was first considered theoretically by Ajiki and Ando, and is hence called Ajiki-Ando (A-A) splitting \cite{A-A}. The observation of A-A splitting in various SWNTs has been reported by many authors by means of the absorption \cite{ZaricScience, Zaric} or by photoluminescence (PL) spectroscopy \cite{Zaric,MortimerMO}. 
Owing to the interplay between inter- and intra-K-K-valley short-range scattering, the exciton states are very complicated with 16 split states of the bright and dark excitons \cite{Zhao}. 
The application of a magnetic field causes the mixing of both states, and it is expected that the complicated exciton states can be clarified experimentally.  
So far, the dark exciton states in ensemble samples have been identified in magneto-PL \cite{MortimerTemp,J.Shaver} or in micromagneto-PL\cite{Matsunaga} in the case of a single SWNT. PL in general is a process arising from a final state after the event of photoexcitation, and is therefore sensitive to unknown impurities or localized states, and is not necessarily a good method for determining and discussing intrinsic and coherent energy states, such as exciton states.
With the aim of observing A-A splitting and the behaviors of these exciton states, we attempted to perform magneto-optical absorption measurements up to an ultrahigh magnetic field.  
The oscillator strength of the optical transition, which can be compared directly with theoretical models, can be derived from the absorption spectra but not necessarily from the PL spectra. We performed magneto-absorption measurements up to a very high magnetic field at which the A-A splitting energy exceeds that of the exciton exchange interaction.\\
\section{Experimental}
 The magnetic fields of up to 78 T were generated by a recently developed giant single-turn coil (GSTC) method\cite{Kindo}. 
Magneto-optical absorption measurements on the chirality specific SWNTs at room temperature were carried out at a near-infrared region.
The SWNTs were grown from alchohol  catalityic CVD (ACCVD), and the SWNTs were dispersed in a liquid, whose synthesis started with a mixture of PFO polymer, {\it d}-toluene, and acetic acid (PFO-ACCVD) \cite{PFO.Murakami}. 
The polymer of 9,9-dioctylfluorenyl-2,7-diyl referred as PFO was used as a solvent for SWNTs \cite{A.Nish}. 
The PFO-SWNTs exhibit a very sharp absorption spectral peak owing to the well-defined chirality of the SWNTs. 
We focused on the absorption spectra from the first inter-subband $E_{11}$ transition exhibiting sharp spectral peaks. 

The spectra at 0 T in Fig.~\ref{fig:Spectra_B} show the absorption of the PFO-ACCVD/{\it d}-toluene samples measured in a glass cell with an inner thickness of about 2 mm. Owing to the significant consequence of the high selectivity of the polymers, the number of different nanotube species is substantially reduced. 
Hence, the spectra are well resolved and the underlying background is significantly reduced (cf. \cite{ZaricScience,Zaric}). 
The full width at half maximum (FWHM) of the well-defined peaks is typically 20 meV, which indicates high spectral sharpness compared with that of SWNTs synthesized by other methods. 
Owing to the high selectivity, 
the spectral peaks of (7,5), (7,6), and (8,6) can thus be regarded as only one chiral type of SWNT, and are focused on in the analysis of A-A splitting expected to be observed in magnetic fields.


The optical transmission spectra in magnetic fields were measured using two different systems. 
The first system is a nondestructive pulse magnet with an inner bore of 20 mm and a pulse width of about 37 ms.  The maximum accessible  magnetic field is about 55 T for the discharge current supplied from a 990 kJ capacitor bank system \cite{Kindo}.  
For fields higher than 55 T, we employed a GSTC system \cite{Kindo} which is capable of generating fields of approximately 100 T in a room-temperature bore of 30 mm with a pulse width of about 80 $\mu$s. 
The operation is destructive since the coil explodes outwardly after the field is generated. 
A maximum electric current of approximately 2-3 MA is injected into the GSTC from a 5 MJ high-speed capacitor bank system. 
The gate of an optical detector consisting of InGaAs diode arrays is synchronously opened at the top of the pulse field , and the exposure time was chosen to maintain the field variation within 3\% during the gate operation  (exposure times: 1.5 ms and 5 $\mu$s, respectively, for the case of the nondestructive pulse magnet and for the case of GSTC). 


A Xe-flash arc lamp is used as a light source. The transmission signals transferred by an optical fiber are dispersed by a 0.2 m grating polychromator. 
The sample cell is mounted in the Voigt configuration with the incident light polarized parallel ($E \parallel B$) or perpendicular ($E \perp B$) to the applied magnetic field ($B$) direction by alternating a linear polarizer attached on the sample cell, and the absorption spectra were taken at the top of each magnetic field.
The liquid sample was held in the restricted space in the bore of the magnet by a hand made miniature glass cell.

\section{Results and Discussion}

\begin{figure}
\includegraphics[width=0.8\columnwidth]{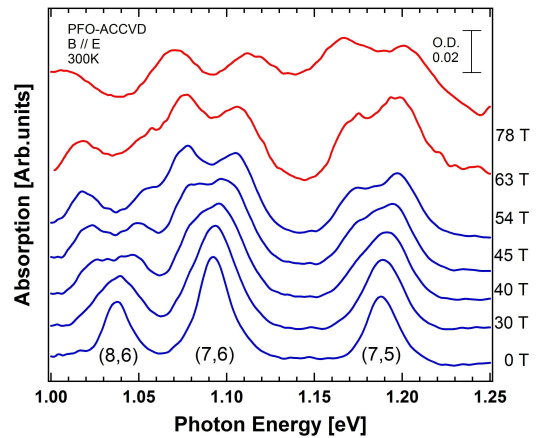}
\caption{\label{fig:Spectra_B} 
Absorption spectra of PFO-ACCVD SWNTs at room temperature measured in pulse magnetic fields.  The $\vec{k}$ $\perp$ $\vec{B}$ ( $\vec{k}$:  light propagation vector) and $\vec{E}$ $\parallel$ $\vec{B}$ (Voigt) configuration. Blue lines are obtained from a nondestructive pulse magnet, and red lines are obtained from the GSTC. O.D. is the optical density of the spectra.
}
\end{figure}

In Fig.~\ref{fig:Spectra_B},  the absorption spectra shown by blue lines are obtained by the nondestructive pulse magnet, which generated a field of up to 54 T, at a temperature of 290 K. 
The distinctive sharp peak for each chirality in the absence of a magnetic field gradually splits  into two peaks upon the application of the external magnetic field. 
It is evident from the spectral evolution that a new peak appeared at the low-energy side of the main peak with increasing magnetic field. 
According to the calculated dynamical conductivity, used by Ando \cite{Ando} to describe the absorption spectrum, this behavior clearly indicates the lower energy location of the dark excitons. 
This result is consistent with those reported by other groups \cite{MortimerMO, MortimerTemp,J.Shaver,Matsunaga,ShaverCrooker,Srivastava}. 
The red lines in Fig.~\ref{fig:Spectra_B} are the spectra measured using the GSCT system of magnetic fields of up to 78 T.
Note that  that each splitting is well separated but that peak broadening becomes more pronounced in fields above 63 T. 

The absorption spectrum for each chirality is deconvoluted by a Gaussian waveform, and the results of the peak shifts and splitting are plotted against the magnetic field. The peak splitting is insufficiently resolved for reliable spectral deconvolution at $B <$ 35 T.  
The obtained peak shifts are plotted against $B$ taken from both the nondestructive long-pulse magnet and the destructive short-pulse GSTC. 
Since the samples are ensembles of SWNTs, two important factors must be considered. One is the magnetic field induced orientation of the SWNTs in an aqueous solution \cite{ZaricScience, Nakamura, JShaver_Orient}. The other is the ensemble average of the magnetic field effectively applied parallel to the tube axis.
The nematic order parameter defined by $S = \frac{1}{2}(3<cos^{2}\theta> - 1)$, is correlated to the optical anisotropy, defined by
 $A = (\alpha_{\parallel} - \alpha_{\perp}) / (\alpha_{\parallel} + 2\alpha_{\perp})$, where $\theta$ is the average angle between the ensemble nanotubes and the direction of $B$, and $\alpha_{\parallel}$ and $\alpha_{\perp}$ are the absorption intensity in the cases of $B \parallel E$ and $B \perp E$, respectively \cite{ Murakami, Nakamura, JShaver_Orient}. 
There is a large difference in $<\theta>$ between the two experiments owing to the three orders of magnitude difference of the pulse rise time of the magnetic field. 

Furthermore, the absorption spectra in Fig.~\ref{fig:Spectra_B} are a result of convoluting the spectra of all randomly oriented nanotubes; the effective $B_{eff}$ should be corrected to take account of  the contribution from all the randomly oriented nanotubes with respect to the mean orientation $<\theta>$ directed by the external magnetic field. Hence $B_{eff} \sim B_{\parallel} \cos 30^\circ$, where $B_{\parallel} = B\cos<\theta>$, and the correction term $\sim \cos30^\circ$ is obtained from a simulation of the spectrum convolution with the assumption that the peak intensity is proportional to sin$\varphi$ when the nanotube axis is tilted by an angle $\varphi$ from the direction of $B$.

\begin{figure}
\includegraphics[width=7cm]{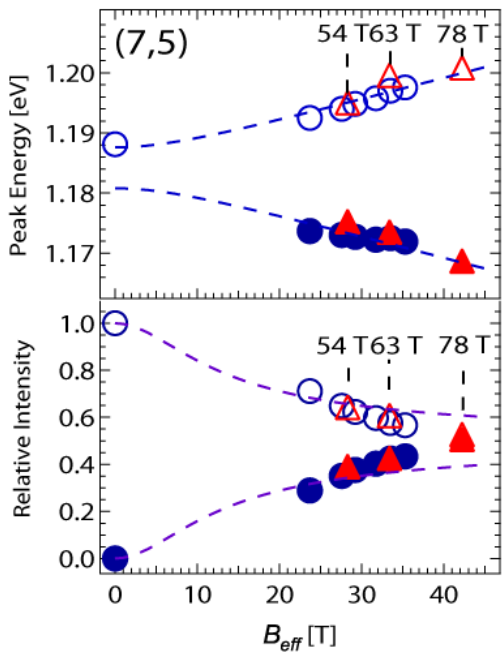}
\caption{ \label{fig:ShiftEI} 
Change in the absorption and the normalized intensity of the spectral peaks for (7,5)-chirality SWNTs plotted against the effective magnetic field $B_{eff}$. Blue marks are results obtained from the nondestructive pulse magnet, and red marks are those obtained from the GSTC.
The dashed lines are the results of fitting.
}
\end{figure}

The peak splitting and normalized absorption intensities are plotted against the effective magnetic field $B_{eff}$ in Fig.~\ref{fig:ShiftEI}. A substantial reduction of $B_{eff}$ can be observed for the data obtained from the GSTC owing to the three orders of magnitude faster rise time of the magnetic field.
The data obtained from two independent experiments are now smoothly connected, and the fitted (dashed) lines consistently reproduce the experimental data up to the highest field.
The peak positions of the split spectra are fitted by the following formula used in the two-level model, similarly to those employed in the previous studies \cite{ J.Shaver, Matsunaga}:
$E^\pm = \pm \sqrt{{\Delta_{bd}}^2+{\Delta_{A-A}}^2}$,
 where $\Delta_{bd}$ represents the bright-dark zero-field splitting and $\Delta_{A-A} = \mu B$ is the so-called AB (or A-A) splitting energy.
 The normalized absorption intensity in Fig.~\ref{fig:ShiftEI} is fitted by the formula
$I^\pm =1/2 \pm \Delta_{bd}/2 \sqrt{{\Delta_{bd}}^2+{\Delta_{A-A}}^2}$.
The parameters $\Delta_{bd}$ and $\mu_{ex}$ were determined so as to fit both sets of data in Fig.~\ref{fig:ShiftEI}.
 A similar procedure is successfully applied to the spectra of (7,6) chirality. However, for (8,6) chirality, the spectral deconvolution becomes ambiguous at high magnetic fields owing to the higher energy component of the split peak merging into the low-energy tail of the (7,6) spectral peak as shown in Fig.~\ref{fig:Spectra_B}.
\begin{center}
\begin{table}[h]
\centering
\begin{tabular}{ccccccc}
\hline
Chirality & $d$&$\mu_{ex}$ &$\mu_{th}$ &ratio& $\Delta_{bd}$&Lower State\\ 
 & [nm]& [meV/T]& [meV/T]&   $\mu_{ex} /\mu_{th}$ &[meV]& \\
\hline
(7,5) & 0.83 & 0.73 & 0.93 &     0.78 &      6.8&    Dark\\
(7,6) & 0.89 & 0.77 & 0.98 &     0.78 &       9.3&     Dark\\
\hline
\end{tabular}
\caption{\label{tab:5/tc}Parameters obtained from the A-A splitting. $d$ = nanotube diameter.}
\end{table}
\end{center}
As a result of the fitting shown in Fig.~\ref{fig:ShiftEI} for both (7,5) and (7,6) chiralities, the values of dark- and bright-exciton splitting $\Delta_{bd}$, and of the coefficient $\mu$ determining the A-A splitting in a magnetic field, given by ${\Delta_{A-A}=\mu B_{eff}}$, are summarized in Table 1, together with the positions of the dark and bright excitons in terms of energy. 
$\mu_{ex}$ is 78\% of $\mu_{th}$, where $\mu_{th}=3 (\pi ed^2/2h) E_{g}$, 
and $E_{g}$ is given by $4 \pi \gamma /3L$,  
$d$ and $L$  the diameter and circumference of the nanotube, respectively, 
and the band parameter $\gamma$ is the same as defined in  \cite{A-A}. 
It is considered that $\mu_{th}$ may be reduced for example by the effect of the magnetic field on the exciton binding energy, although this is still insufficient to explain the overestimation relative to the present experimental values.  The present data suggest that the simple model must be modified.
\begin{figure}
\includegraphics[width=0.8\columnwidth]{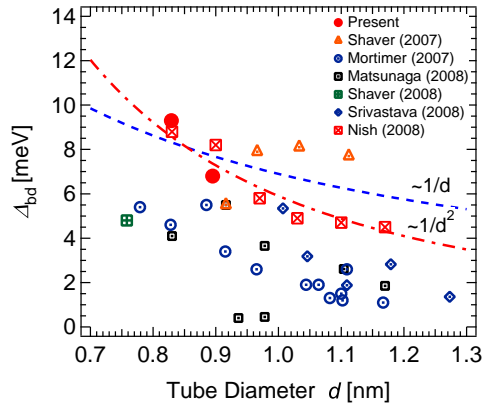}
\caption{\label{fig:Delta_bd} 
Bright-dark exchange splitting $\Delta_{bd}$ plotted against the tube diameter. The present data are compared with those from various previous works described in detail in the text.
}
\end{figure}

The values of bright-dark exchange splitting $\Delta_{bd}$ are larger than those reported in previous reports \cite{MortimerTemp,J.Shaver,ShaverCrooker,Matsunaga,Srivastava}. 
$\Delta_{bd}$ is plotted against the tube diameter $d$ together with the data obtained by other groups in Fig.~\ref{fig:Delta_bd}. 
The present data agree reasonably well with some of those obtained by Shaver {\it et al.}\cite{J.Shaver}, by Srivastava {\it et al.} \cite{Srivastava} and by Nish {\it et al.} \cite{Nish_delta}. 
The splitting in magnetic fields is assumed to be proportional to $1/d$ which is predicted by tight-binding theory (red dashed line) \cite{Perebeinos} or to $1/d^{2}$ according to Spataru {\it et al.} by first-principles theory (blue dot-dashed line) \cite{Spataru}. 
In 2007, Shaver {\it et al.} used magnetic-field-induced PL brightening in a stretch-aligned gelatin film containing individual SWNTs up to a  high pulsed magnetic field of 56 T \cite{J.Shaver}.  
All the other low-lying data below 6 meV in Fig.~\ref{fig:Delta_bd} were obtained by PL in rather low magnetic fields except for the data obtained by Shaver {\it et al.} in 2008 \cite{ShaverCrooker}, who used a low-energy shift of a PL peak in magnetic fields to evaluate $\Delta_{bd}$ as a result of fitting in a  magnetic field of 30-50 T. 
It is plausible that their value of  $\Delta_{bd}$ is underestimated, since in the range of magnetic fields, the peak shifts are still affected by the mixing of dark and bright states (see Fig. 4 in \cite{ShaverCrooker}). 

Matsunaga {\it et al.} performed a micro-PL measurements of samples that were spatially isolated \cite{Matsunaga}. 
They evaluated the values obtained from PL peak splitting in magnetic fields, which should cause the energy shift symmetrically in the high- and low-energy directions, using the same equation as us. Their high-energy peak, however, exhibited almost constant values in magnetic fields. There is clearly a contradiction between the behavior of the data and the formula used. 
Their values approximately coincide with ours after they are increased twofold. 
Mortimer and Nicholas  considered the temperature dependence of PL in the absence and presence of a magnetic field (19.5 T), and evaluated the splitting using a model of the PL thermal population \cite{MortimerTemp}.  
They observed zero-field splitting due to stress induced by the thermal expansion of ambient materials. 
As far as the effect of stress is comparable to that of an external magnetic field, their values of $\Delta_{bd}$ could be associated with a systematic error and may not merit quantitative discussion.
Some of the values reported by Srivastava {\it et al.} \cite{Srivastava} are also small. 
They conducted micro-PL measurements on individual HiPco SWNTs on quartz substrates. 
Owing to the sharpness of the PL spectrum, they could determine the precise positions of the PL peak splitting. Again, the high-energy PL peak behaved in a strange manner, shifting to the lower energy side. Therefore, the application of the equation described above is still contradictory, similarly to the case of Matsunaga {\it et al.} as mentioned above.

The results obtained by  by Jiang {\it et al}.  \cite{Jiang} and Capaz {\it et al}. \cite{Capaz} employing  the tight binding approach are reasonably close to our values of $\Delta_{bd}$, as well as part of the data reported by Shaver {\it et al.}  \cite{ShaverCrooker}. 
On the other hand, the theory of {\it ab initio} calculation employed by Spataru {\it et al.} \cite{Spataru} leading to a value of approximately 30 meV still seems to overestimate $\Delta_{bd}$. 
The physical reason for the $1/d^{2}$ dependence is explained by a long-range contribution to the exchange energy according to the discussion in Capaz {\it et al.}\cite{Capaz}. However, as can be seen in Fig.~\ref{fig:Delta_bd}, owing to limited sample size of the present data, it is difficult to conclude whether the tube diameter dependence of $\Delta_{bd}$ is $1/d$ or $1/d^2$. However, present data together with those obtained by some of the previous authors favor the  $1/d^2$ dependence.

The A-A splitting in SWNTs was first reported by Zaric {\it et al.} \cite{ZaricScience} as a result of magneto-absorption and PL measurements in magnetic fields of up to 45 T, but owing to the unresolved splitting of the absorption peaks of ensemble nanotubes, they were unable to determine a precise value for $\mu_{ex}$. 
The A-A splitting measured up to 74 T was well described by the model simulation, assuming an average splitting of 0.7-0.9 meV/T for samples with various chiralities overlapping each other in their spectra \cite{Zaric}, which agrees reasonably well with the present data. 
Mortimer {\it et al.} \cite{MortimerMO} obtained the PL energy shifts due to A-A splitting in SWNT solutions with different tube diameters at room temperature  in magnetic fields up to 58 T. They claimed that the A-A shifts follow  quantitative predictions with values being approximately 70\% of those predicted for completely oriented nanotubes. 
However, the experimental shifts were slightly larger than those expected after a 50\% correction was applied to take account of the random orientation of the ensemble nanotubes.
Note that the present value of $\mu_{ex}$ is 0.78 times that of the calculated value for both chiralities in this study \cite{A-A}.

\section{Summary}
In summary, the well resolved exciton A-A splitting was obtained from the high-field magneto-absorption spectra of  the $E_{11}$ transitions in PFO-SWNTs. The A-A spectral peak splitting and the exciton exchange energies were determined for samples with (7,5) and (7,6) chiralities. The band-edge dark excitons were found to be located at a lower energy than the bright excitons.
The value of $\mu_{ex}$ is almost 80\% of that of $\mu_{th}$ obtained from theory.

\end{document}